# A Simpler Explanation to BAK1 Gene Variation in Aortic and Blood Tissues

Michel E. Beleza Yamagishi[1]

Laboratório de Bioinformática Aplicada, Embrapa Informática Agropecuária, 209,Andre Tosello Av. Campinas-SP, Brazil

We read with great interest the article written by Gottlieb et al. [2009] where they report some intriguing polymorphism differences in aortic and blood tissues in the same individual. Actually, they were interested to perform case-control association analysis in AAA patients; however, instead of obtain a significant association, they realized that both diseased and non-diseased AA tissues showed the same polymorphisms which was not observed in the matching blood samples. This observation was not expected, and raised many speculations about its meaning and it has received wide press coverage. The most controversial speculation says that this observation jeopardize the current paradigm that cells from the same individual have the same genome. Actually, in a first approach, one is led to either review the paradigm or find an alternative explanation that reconciles the experiments and the current paradigm.

We have an alternative explanation that seems to be correct for two reasons: (i) it is simpler and (ii) it explains the BAK1 gene variations in different tissues. The explanation is that in aortic tissue (both diseased and nondiseased) a BAK1 pseudogene is expressed; while in the matching blood samples the actual BAK1 gene is expressed. This explanation was reached after we realized that BAK1 has two edited copies in human genome. These copies are probably BAK1 pseudogenes. One copy belongs to chromosome 11 (NG_005599.3) and the other to chromosome 20 (NC_000850.5). The first copy has frameshifts which means that probably it does not express any functional protein; by other hand, the chromosome 20 copy has no frameshifts and what is more important contains all the reported polymorphisms. It is known that some pseudogenes are actually transcribed [Balakirev & Ayala, 2003], so it is not unreasonable to speculate that BAK1 pseudogene on chromosome 20 is transcribed in aortic tissue.

---

[1] Correspondence to: michel@cnptia.embrapa.br

Although, we decided to use sequence NM_001188 as BAK1 gene reference sequence, we noted that there are some differences between its sequence and Gottlieb and coauthors reference sequence. For instance, the 28[th] amino acid in Gottlieb's sequence is ALA which is coded by GTC codon, while in sequence NM_001188 the codon is GCC. See table 1 for more differences. Furthermore, there are also some inconsistencies, probably copy editing errors as pointed out by Forsdyke in a personal communication, for example, GCT codes ALA instead of VAL as appears in Gottlieb et al. table 2. Observe that the 42[nd] amino acid is supposed to code HIS but its codon is CAA in their table 2 which in fact codes for GLN.

Table 1: Codons differences (our corrections are written in red). The reported polymorphisms between BAK1 gene and BAK1 pseudogene on chromosome 20 are also shown.

| Amino acid # | 28 | 42 | 52 | 81 | 103 |
|---|---|---|---|---|---|
| Gottlieb's refseq | GTC (ALA) | CAG (ARG) | GCT (ALA) | ATC (ILE) | GCC (ALA) |
| BAK1 gene | GCC (ALA) | CGC (ARG) | GTG (VAL) | ATC (ILE) | ACG (THR) |
| BAK1 pseudogene | GCC (ALA) | CAC (HIS) | GCG (ALA) | ATT (ILE) | ACG (THR) |

## *Acknowledgements*

We thank José Andres Yunes who initially drew the matter to our attention and to Donald Forsdyke for noting the errors in Table 2 of Gottlieb et al. and for advice in the preparation of this letter.